\documentclass[conference]{IEEEtran}
\IEEEoverridecommandlockouts
\usepackage{cite}
\usepackage{amsmath,amssymb,amsfonts}
\usepackage{algorithmic}
\usepackage{graphicx}
\usepackage{textcomp}
\usepackage{xcolor}
\usepackage{booktabs}
\usepackage{multirow}
\usepackage{subfig}
\usepackage{algorithmic}
\usepackage{algorithm}
\def\BibTeX{{\rm B\kern-.05em{\sc i\kern-.025em b}\kern-.08em
    T\kern-.1667em\lower.7ex\hbox{E}\kern-.125emX}}
\begin{document}

\title{Federated Breast Cancer Detection Enhanced by Synthetic Ultrasound Image Augmentation\\
\thanks{This work was supported by the following NIH grants: NCI R01-CA246704, R01-CA240639, U01-CA268808, and NHLBI R01-HL171376.}
}

\makeatletter
\newcommand{\linebreakand}{%
  \end{@IEEEauthorhalign}
  \hfill\mbox{}\par
  \mbox{}\hfill\begin{@IEEEauthorhalign}
}
\makeatother

\author{\IEEEauthorblockN{Hongyi Pan}
\IEEEauthorblockA{\textit{Department of Radiology} \\
\textit{Northwestern University}\\
Chicago, IL, USA \\
hongyi.pan@northwestern.edu}
\and
\IEEEauthorblockN{Ziliang Hong}
\IEEEauthorblockA{\textit{Department of Radiology} \\
\textit{Northwestern University}\\
Chicago, IL, USA \\
ZiliangHong2029@u.northwestern.edu}
\and
\IEEEauthorblockN{Gorkem Durak}
\IEEEauthorblockA{\textit{Department of Radiology} \\
\textit{Northwestern University}\\
Chicago, IL, USA \\
gorkem.durak@northwestern.edu}
\linebreakand 
\IEEEauthorblockN{Ziyue Xu}
\IEEEauthorblockA{
\textit{NVIDIA}\\
Bethesda, MD, USA \\
ziyuex@nvidia.com}
\and
\IEEEauthorblockN{Ulas Bagci}
\IEEEauthorblockA{\textit{Department of Radiology} \\
\textit{Northwestern University}\\
Chicago, IL, USA \\
ulas.bagci@northwestern.edu}
}

\maketitle

\begin{abstract}
Federated learning enables collaborative training of deep learning models across institutions without sharing sensitive patient data. However, its performance is often limited by small datasets and non-independent, identically distributed data, which can impair model generalization. In this work, we propose a generative model-based data augmentation framework for breast ultrasound classification. It leverages synthetic images generated by deep convolutional generative adversarial networks and a class-conditioned denoising diffusion probabilistic model. Experiments on three publicly available datasets (BUSI, BUS-BRA, and UDIAT) demonstrated that incorporating a suitable number of synthetic images improved average AUC from 0.9206 to 0.9362 for FedAvg and from 0.9429 to 0.9574 for FedProx. Furthermore, we noticed that excessive use of synthetic data reduced performance. This highlights the importance of balancing real and synthetic samples. Our results underscore the potential of generative model-based augmentation to enhance federated breast ultrasound image classification.
\end{abstract}


Breast cancer is the most prevalent type of cancer and one of the leading causes of cancer-related deaths among women. Recently, deep learning techniques have achieved remarkable success in automating breast ultrasound image classification~\cite{daoud2019breast,taheri2025improving,qezelbash2025hybrid} and segmentation~\cite{he2023hctnet,zhang2025fet,wang2025cross} tasks. However, training robust and generalizable models typically requires large-scale, diverse datasets that are often challenging to obtain due to privacy regulations and data-sharing restrictions across medical institutions.

Generative Adversarial Networks (GANs)~\cite{goodfellow2014generative} have been widely explored for medical image synthesis, particularly in scenarios where annotated datasets are scarce and difficult to acquire. By training a generator and discriminator in an adversarial manner, GANs can produce visually realistic samples that enrich training data and improve model generalization. The original “vanilla” GAN architecture inspired the development of numerous variants that aimed to address its training instability and limited image fidelity. Among these, Deep Convolutional GANs (DCGANs)~\cite{radford2015unsupervised,medghalchi2025synthetic} introduced convolutional and transposed convolutional layers, which provided more stable training dynamics and sharper outputs. As a result, DCGANs became a practical and widely adopted choice for medical image augmentation across different imaging modalities. Building on these advances, StyleGAN~\cite{karras2019style} further improved both image fidelity and controllability by incorporating a style-based generator, making it possible to synthesize high-resolution images with fine-grained control over visual attributes. This level of flexibility enabled more realistic augmentation scenarios in medical applications, particularly when capturing subtle variations in anatomy or pathology. More recently, diffusion probabilistic models (DDPMs)~\cite{ho2020denoising,rombach2022high,zhang2023adding,medghalchi2025synthetic} have emerged as a compelling alternative to GAN-based approaches. By modeling image generation as a gradual denoising process, DDPMs avoid some of the instability issues inherent in adversarial training and produce samples with higher diversity and stability. These properties make diffusion models especially attractive for medical imaging, where reliable and diverse synthetic data can help mitigate class imbalance and improve model robustness.

Federated learning enables collaborative model training across multiple institutions while preserving data privacy by avoiding direct data sharing~\cite{mcmahan2017communication,li2020federated,li2021fedbn}. Challenges such as non-IID data distribution, domain shifts, and limited local data hamper federated learning performance. Recent works have explored personalized federated learning~\cite{tan2022towards} and domain generalization~\cite{liu2021feddg,zhang2023federated,pan2024domain,pan2025frequency} to mitigate these issues. 


In this work, we propose a federated learning framework that takes synthetic breast ultrasound images from two class-specific DCGANs~\cite{radford2015unsupervised} and a class-conditioned DDPM~\cite{ho2020denoising} for data augmentation across clients. Experiments on three publicly available datasets (BUS-BRA~\cite{gomez2024bus}, BUSI~\cite{al2020dataset}, and UDIAT~\cite{yap2017automated}) demonstrated that introducing a suitable amount of synthetic images consistently improves classification performance. Our ablation study further revealed that excessive synthetic data can degrade both accuracy and AUC. This finding underscores the importance of carefully balancing real and synthetic samples to enhance model generalization while preserving data privacy.

\section{Methodology}

\subsection{Problem Statement}

Given breast ultrasound datasets distributed across multiple clinical centers, our goal is to collaboratively train a binary classification model to distinguish benign from malignant lesions under the federated learning setting. Let there be $K$ clients, each with a local dataset $\mathcal{D}_K=\{(x_0^k, y_0^k), \ldots, (x_{n_k-1}^k, y_{n_k-1}^k)\}$, where
$x_i^k$ is a BUS image and $y_i^k$ is the corresponding label (0 for benign, 1 for malignant). In standard FL, the global model $f_\theta$ is optimized by minimizing the weighted sum of local losses without exchanging raw data:
\begin{equation}
\min_{\theta} \sum_{k=0}^{K-1} \frac{n_k}{n} \mathcal{L}_k(\theta), 
\end{equation}
where
\begin{equation} \mathcal{L}_k(\theta) = \frac{1}{n_k} \sum_{i=0}^{n_k-1} \ell(f\theta(x_i^k), y_i^k).
\end{equation}
Here, $\ell(\cdot)$ denotes the binary cross-entropy loss, and $n=\sum_{k=1}^{K} n_k$ is the total number of samples. However, this setup faces practical challenges such as non-IID data distributions across clients, limited local data (especially malignant cases), and poor generalization. To overcome these limitations, we propose augmenting each client’s training set with synthetic BUS images from generative models. These synthetic images $\tilde{\mathbf{x}}\sim G_c(z)$, where $G_c(\cdot)$ is the generator for class $c\in\{0, 1\}$ and $z \sim \mathcal{N}(0, \mathbf{I})$ are distributed equally to all clients to enrich data diversity, alleviate class imbalance, and improve federated model convergence and robustness without compromising data privacy.

\subsection{Synthetic breast ultrasound images generation via DCGAN}
To address the challenges of data scarcity and non-IID distributions in federated breast ultrasound classification, we incorporate synthetic image generation using DCGANs~\cite{radford2015unsupervised}. GAN-based data augmentation has shown promise in medical imaging for enriching training sets with realistic, domain-relevant variations~\cite{irmakci2022multi}. In this work, we do not use diffusion models because, despite their recent success in producing high-quality and diverse images, they are computationally intensive and exhibit slow sampling speeds. Specifically, we train two separate DCGAN models: one for generating benign images and one for malignant images. Both models are trained centrally using the combined training and validation datasets while strictly excluding all test data to prevent data leakage. The generator network maps a random noise vector concatenated with a lesion mask to a grayscale BUS image, aiming to produce realistic ultrasound textures while maintaining anatomical plausibility. The discriminator learns to differentiate between real and synthetic images, enforcing the realism of generated outputs.

Formally, let $z \sim \mathcal{N}(0, I)$ be the latent noise vector and $M$ be a binary lesion mask. The generator $G(z, M)$ outputs a synthetic ultrasound image $I_{syn}$, while the discriminator $D$ attempts to distinguish real images $I_{real}$ from $I_{syn}$. The objective functions for $G$ and $D$ follow the standard DCGAN formulation:
\begin{equation}
\min_G \max_D \; \mathbb{E}_{\mathbf{x} \sim p_{\text{data}}(\mathbf{x})} [\log D(\mathbf{x})] + 
\mathbb{E}_{\mathbf{z} \sim p_{\mathbf{z}}(\mathbf{z})} [\log (1 - D(G(\mathbf{z})))].
\end{equation}

Fig.~\ref{fig: GAN} illustrates the architecture of the DCGAN employed in this study. The generator network produces a $128\times 128$ synthetic breast ultrasound image from a 128-dimensional latent noise vector using six 2D transposed convolutional layers. The discriminator, composed of six 2D convolutional layers, classifies input images as real or synthetic. We also experimented with generating $256\times256$ images using a similarly scaled architecture, but observed unstable training and poor convergence. As a result, we generate images at a resolution of $128\times 128$, which are subsequently upsampled to $224\times 224$ before being used for training the classification networks.

\begin{figure}[tb]
    \centering
    \subfloat[The generator structure.]{
    \includegraphics[page=2, width=0.9\linewidth]{figures/GAN.pdf}
    }\\
    \subfloat[The discriminator structure.]{
    \includegraphics[page=1, width=0.9\linewidth]{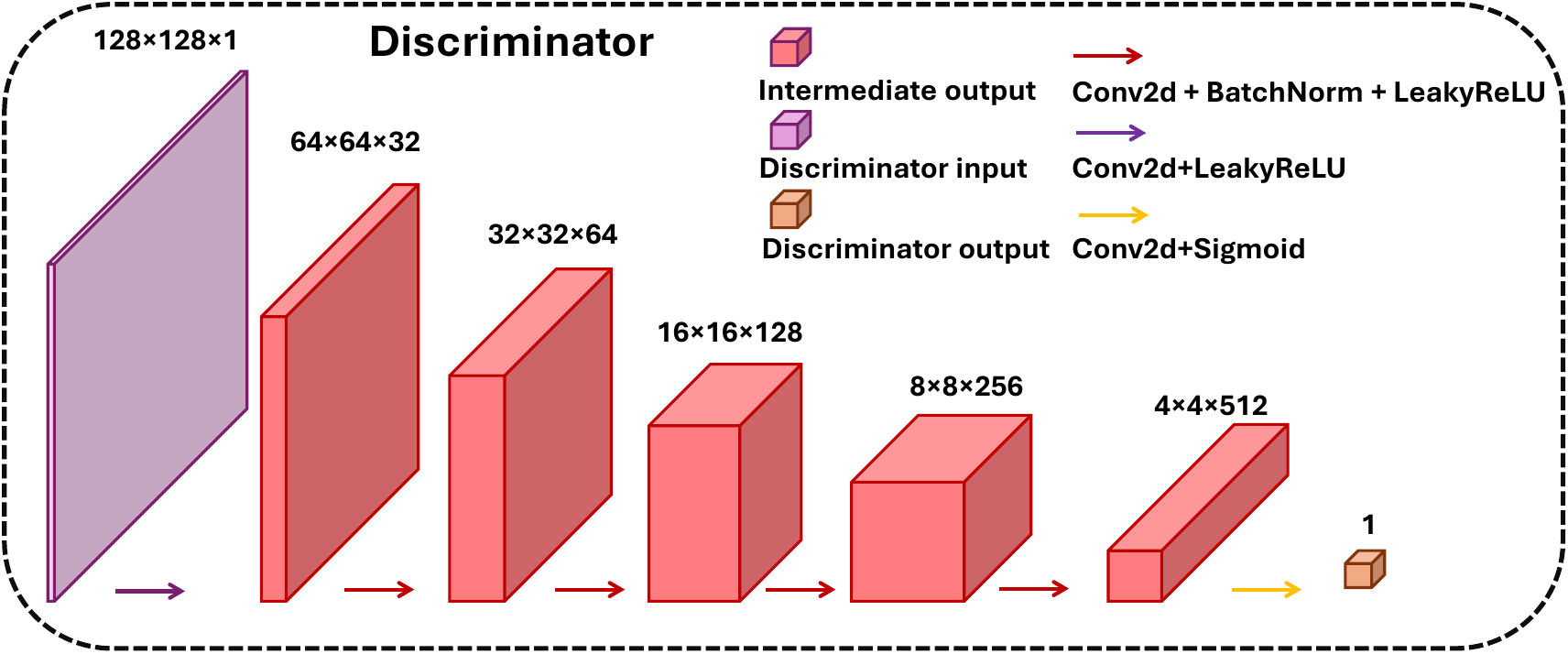}
    }
    \caption{DCGAN for synthetic breast ultrasound image generation.}
    \label{fig: GAN}
\end{figure}

\subsection{Synthetic breast ultrasound image generation via Class-Conditioned Diffusion Model}

To improve the realism and diversity of synthetic breast ultrasound (BUS) images used for federated augmentation, we trained a class-conditioned denoising diffusion probabilistic model (DDPM)~\cite{ho2020denoising}. Rather than training separate generators for each class, a single class-conditioned model was trained. Class-conditioning was implemented via classifier-free guidance: during training, class labels \(y\in\{0,1\}\) (0 for benign and 1 for malignant) were provided to the denoiser network \(\epsilon_\theta(\mathbf{x}_t,t)\), with labels randomly dropped at a fixed probability to enable guidance at sampling time.
\begin{figure}[tb]
    \centering
    \includegraphics[width=1\linewidth]{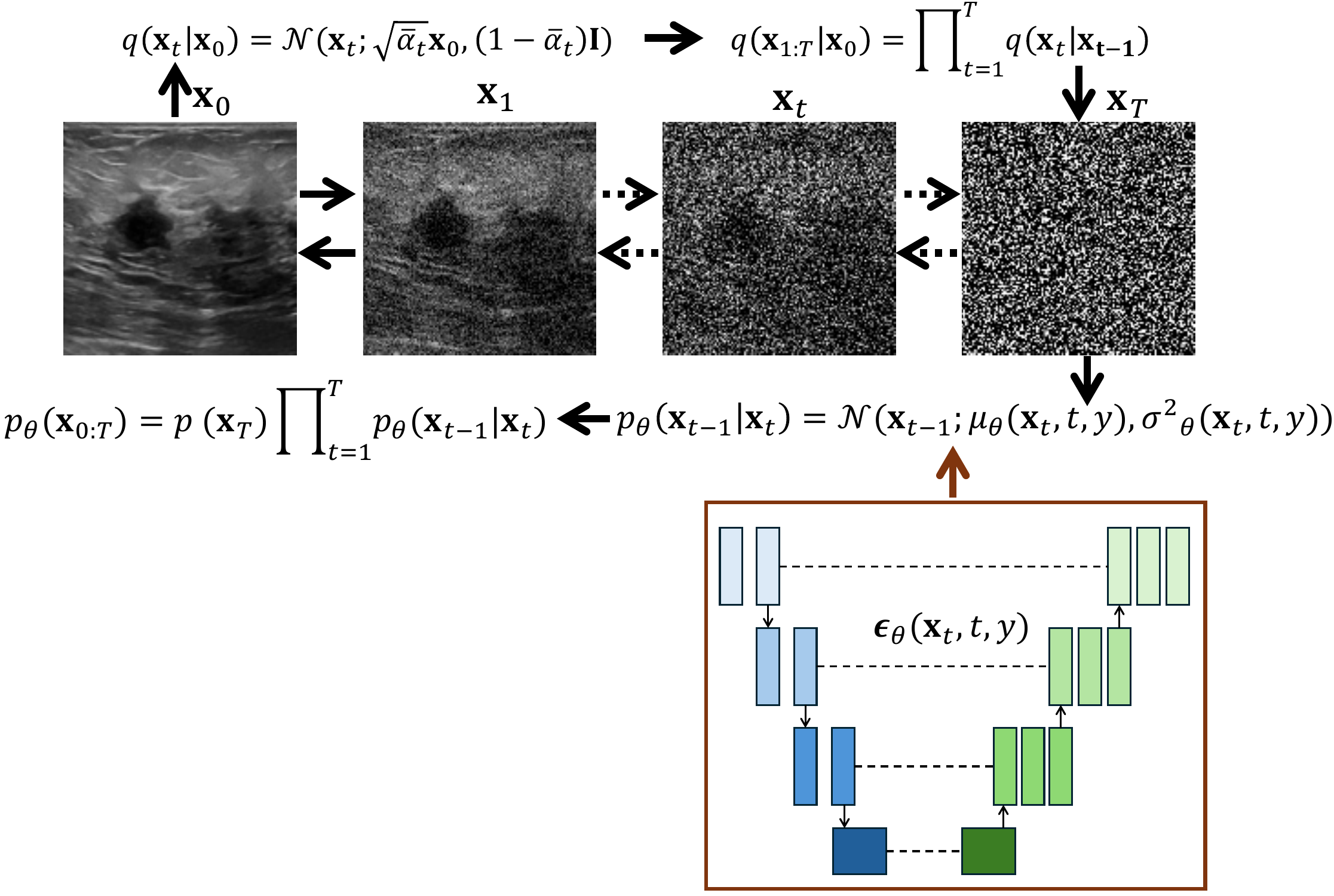}
    \caption{Class-conditioned diffusion model for breast ultrasound image generation.} 
    \label{fig:DDPM}
\end{figure}

As Fig.~\ref{fig:DDPM} shows, the forward (diffusion) process gradually perturbs a clean image \(x_0\) into Gaussian noise:
\begin{equation}
q(\mathbf{x}_t \mid \mathbf{x}_0) = \mathcal{N}\big(\mathbf{x}_t; \sqrt{\bar{\alpha}_t}\,\mathbf{x}_0,\; (1-\bar{\alpha}_t)\mathbf{I}\big), \quad
\bar{\alpha}_t = \prod_{s=1}^t (1-\beta_s),
\end{equation}
with a prescribed noise schedule \(\{\beta_t\}_{t=1}^T\).  
The corresponding reverse (denoising) process is modeled as:
\begin{equation}
p_\theta(\mathbf{x}_{t-1}\mid \mathbf{x}_t,y) = \mathcal{N}\big(\mathbf{x}_{t-1}; \mu_\theta(\mathbf{x}_t,t,y), \, \sigma_t^2\theta (\mathbf{x}_t, t, y)\big),
\end{equation}
where the mean \(\mu_\theta(\mathbf{x}_t,t)\) and the standard deviation \(\sigma_t^2\) are parameterized using the predicted noise \(\epsilon_\theta(\mathbf{x}_t,t,y)\). This defines a Markov chain that gradually denoises the sample from pure noise \(\mathbf{x}_T\sim \mathcal{N}(0,\mathbf{I})\) back to a data-like image \(\mathbf{x}_0\).  

The denoiser is trained with the simplified DDPM objective:
\begin{equation}
\mathcal{L}_{\text{DDPM}}(\theta) \;=\; \mathbb{E}_{\mathbf{x}_0,y,\epsilon,t}\Big[\big\| \epsilon - \epsilon_\theta(\mathbf{x}_t,t,y)\big\|^2\Big],
\end{equation}
where \(\mathbf{x}_t=\sqrt{\bar{\alpha}_t}\,x_0 + \sqrt{1-\bar{\alpha}_t}\,\epsilon\) and \(\epsilon\sim\mathcal{N}(0,\mathbf{I})\).

At inference time, generation begins from pure Gaussian noise $\mathbf{x}_T \sim \mathcal{N}(0, I)$ and applies the reverse process step-by-step until reaching $\mathbf{x}_0$, a synthetic ultrasound image. Class-conditioning ensures that generated samples align with the target lesion category. We generate $256\times256$ synthetic images then downsample them to $224\times 224$ to feed the classification networks for training.

\subsection{Federated Learning with Synthetic Images}
\begin{figure}[tb]
    \centering
    \includegraphics[width=1\linewidth]{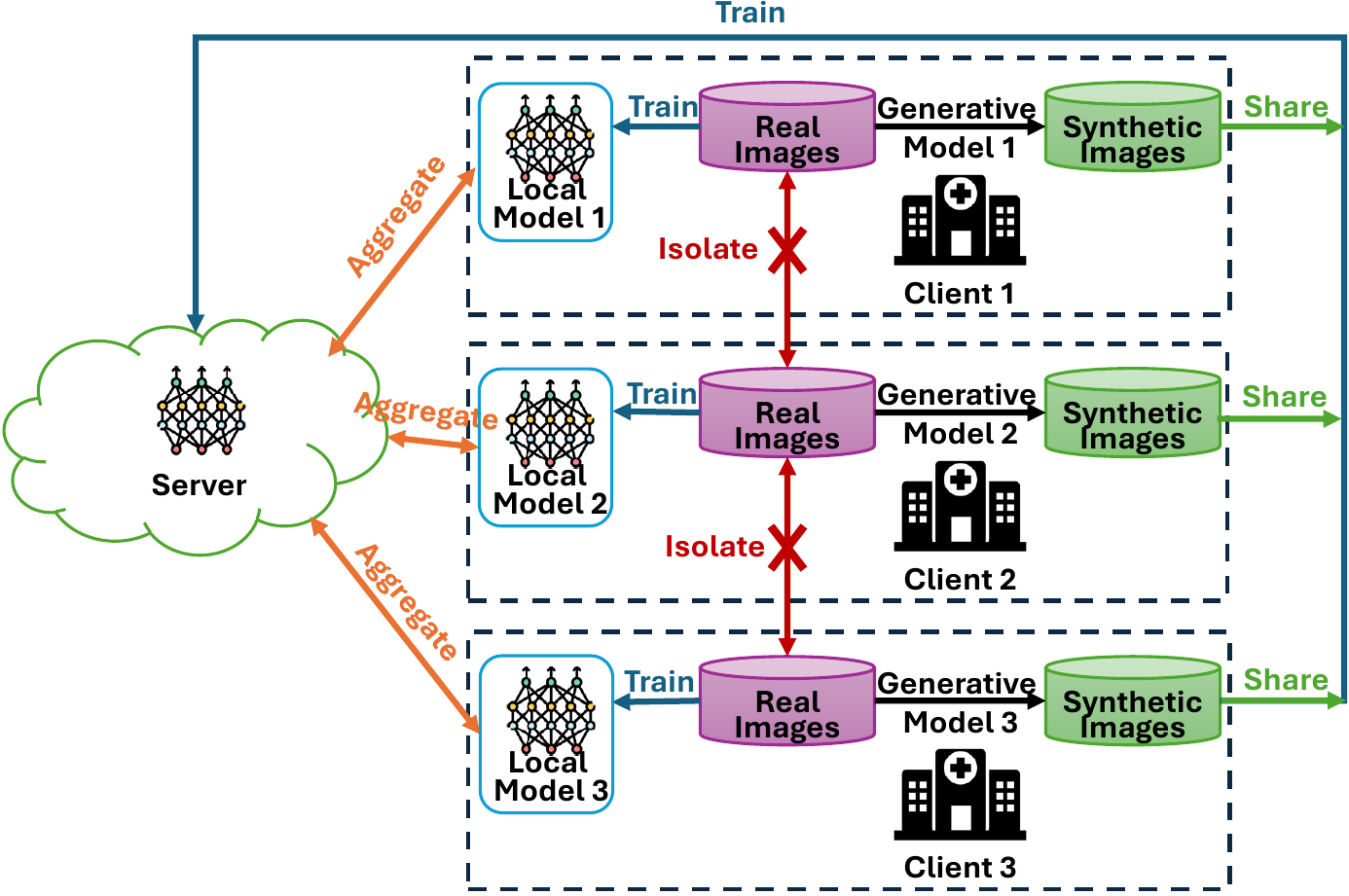}
    \caption{Federated learning framework with synthetic image sharing. Each client locally trains a generative model (DCGAN or DDPM) using its private real images, which remain isolated and are never shared. The trained generative models generate synthetic images that are exchanged among clients and the server. These shared synthetic images are then used to collaboratively train local classification models, enabling multi-center learning while preserving data privacy.} 
    \label{fig:framework}
\end{figure}

To evaluate the utility of synthetic breast ultrasound images in privacy-preserving distributed training, we simulated a federated learning setup where synthetic images were shared across multiple clients. Fig.~\ref{fig:framework} presents the overall federated learning framework. 
Fig.~\ref{fig:framework} presents the overall federated learning framework with synthetic image sharing. In this framework, each client operates on locally stored real images that are strictly isolated and cannot be transmitted due to privacy constraints. Instead of sharing raw data, each client independently trains a generative model, such as DCGAN or DDPM, using its private dataset. Once trained, these generative models produce synthetic images that capture the underlying data distribution without exposing sensitive patient information. The generated synthetic images are then shared and utilized for collaborative training of downstream classification models across clients and the central server. By replacing real data exchange with synthetic image sharing, the proposed framework enables effective multi-center federated learning while maintaining data privacy and institutional data sovereignty.


In this work, we adopted two widely-used federated learning algorithms as baselines: FedAvg~\cite{mcmahan2017communication} and FedProx~\cite{li2020federated}. Both algorithms operate under the standard federated learning protocol: at each communication round, a global model is distributed to all clients, which perform local updates using their data, followed by aggregation at the server. In detail:
FedAvg~\cite{mcmahan2017communication} performs simple weighted averaging of the local model updates:
    \begin{equation}
        \mathbf{w}^{t+1} = \sum_{k=0}^{K-1} \frac{n_k}{n} \mathbf{w}_k^t,
    \end{equation}
    where $w_k^t$ is the model from client $k$, $n_k$ is the number of samples at client $k$, and $n=\sum_{k=0}^{K-1}n_k$.
    FedProx~\cite{li2020federated} introduces a proximal term to stabilize training across heterogeneous data:
    \begin{equation}
        \min_{\mathbf{w}} \; f_k(\mathbf{w}) + \frac{\mu}{2} \| \mathbf{w} - \mathbf{w}^t \|^2,
    \end{equation}
    where $\mu$ controls the strength of the proximal regularization toward the global model $\mathbf{w}^t$.
    
In our study, each client performs multiple local updates using a mix of real and synthetic breast ultrasound images. The synthetic data provides regularization and prevents overfitting to local client-specific distributions, thereby improving global model generalizability. We evaluated both centralized and federated settings to quantify the benefit of incorporating synthetic data in federated breast ultrasound classification.

\begin{algorithm}[tb]
	\caption{Federated learning with synthetic images.}
	\label{al: fed}
	\begin{algorithmic}[1]
		\renewcommand{\algorithmicrequire}{\textbf{Input:}}
		\renewcommand{\algorithmicensure}{\textbf{Output:}}
		\REQUIRE Global model initial wights $\mathbf{w}^0$, training datasets $\mathcal{D}_K$ with the corresponding length $n_k$ for $k=0, 1, \dots, K-1$, synthetic  datasets $\mathcal{S}$. \\\textbf{Hyperparameters:} number of epochs $T$, adaptive aggregation weights step size $s$.
		\ENSURE  Well-trained global model wights $\mathbf{w}^T$.
        \STATE $n=\sum_{k=0}^{K-1}n_k;$
		\FOR{$t=0, 1, \dots, T-1$} 
            \STATE Update $\mathbf{w}^t$ on $\mathcal{S}$;
            \FOR{$k=0, 1, \dots, K-1$} 
            \STATE Assign weights to local model: $\mathbf{w}_k^t=\mathbf{w}^t$;
            \STATE Update $\mathbf{w}_k^t$ using $\mathcal{D}_k$;
            \ENDFOR 
             \STATE Aggregate models: $\mathbf{w}^{t+1} = \sum_{k=0}^{K-1} \frac{n_k}{n} \mathbf{w}_k^t,$;
            \ENDFOR \\
	\end{algorithmic}
\end{algorithm}

\section{Experiments}
\subsection{Dataset Description}

To conduct federated learning experiments for binary classification of benign versus malignant lesions, we leveraged three publicly available breast ultrasound datasets: BUS-BRA~\cite{gomez2024bus}, BUSI~\cite{al2020dataset}, and UDIAT~\cite{yap2017automated}.
Samples of the breast ultrasound images are visualized in Fig~\ref{fig: samples}, and the characteristics of the datasets are summarized in Table~\ref{tab: dataset}. For each dataset, we divided the images into training, validation, and testing subsets using an approximate ratio of 3:1:2, ensuring that both benign and malignant cases were proportionally represented across splits. The training and validation subsets were used for the classification model and the generative models training. Importantly, no testing images were used to train the generative models for preserving the integrity of the evaluation and avoiding data leakage.

\begin{table}[tb]
    \caption{Dataset Description.}
    \centering
    \begin{tabular}{lccc}
    \toprule
         Dataset&Benign&Malignant&Total  \\
         \midrule
         BUS-BRA&1268&607&1875\\
         BUSI&437&210&647\\ 
         UDIAT&109&54&163\\\bottomrule
    \end{tabular}
    \label{tab: dataset}
\end{table}

\begin{figure}[tb]
    \centering
    \subfloat[BUSI benign.]{
    \includegraphics[width=0.3\linewidth, height=0.3\linewidth]{figures/samples/BUSIBenign.png}
    }
    \subfloat[BUS-BRA benign.]{
    \includegraphics[width=0.3\linewidth, height=0.3\linewidth]{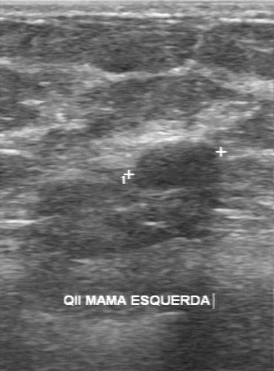}
    }
    \subfloat[UDIAT benign.]{
    \includegraphics[width=0.3\linewidth, height=0.3\linewidth]{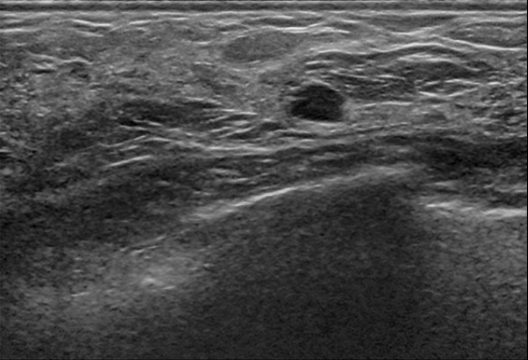}
    }\\
    \subfloat[BUSI malignant.]{
        \includegraphics[width=0.3\linewidth, height=0.3\linewidth]{figures/samples/BUSIMalignant.png}
    }
    \subfloat[BUS-BRA malignant.]{
        \includegraphics[width=0.3\linewidth, height=0.3\linewidth]{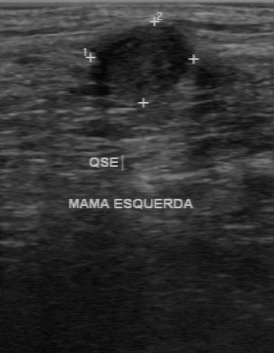}
    }
    \subfloat[UDIAT malignant.]{
        \includegraphics[width=0.3\linewidth, height=0.3\linewidth]{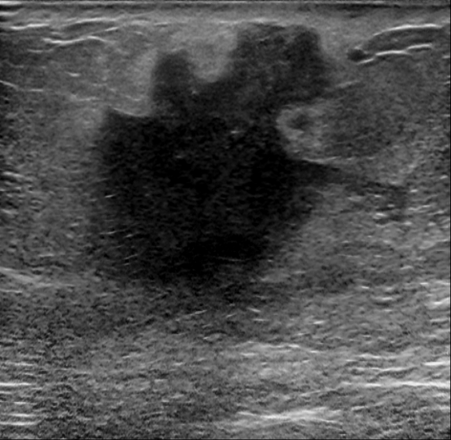}
    }
    \\
    \caption{Breast ultrasound image samples.}
    \label{fig: samples}
\end{figure}


\begin{figure}[tb]
    \centering
    \subfloat[DCGAN benign.]{
    \includegraphics[width=0.3\linewidth, height=0.3\linewidth]{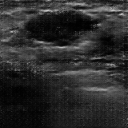}
    }
    \subfloat[DCGAN Malignant.]{
        \includegraphics[width=0.3\linewidth, height=0.3\linewidth]{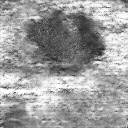}
    }\\
    \subfloat[DDPM benign.]{
    \includegraphics[width=0.3\linewidth, height=0.3\linewidth]{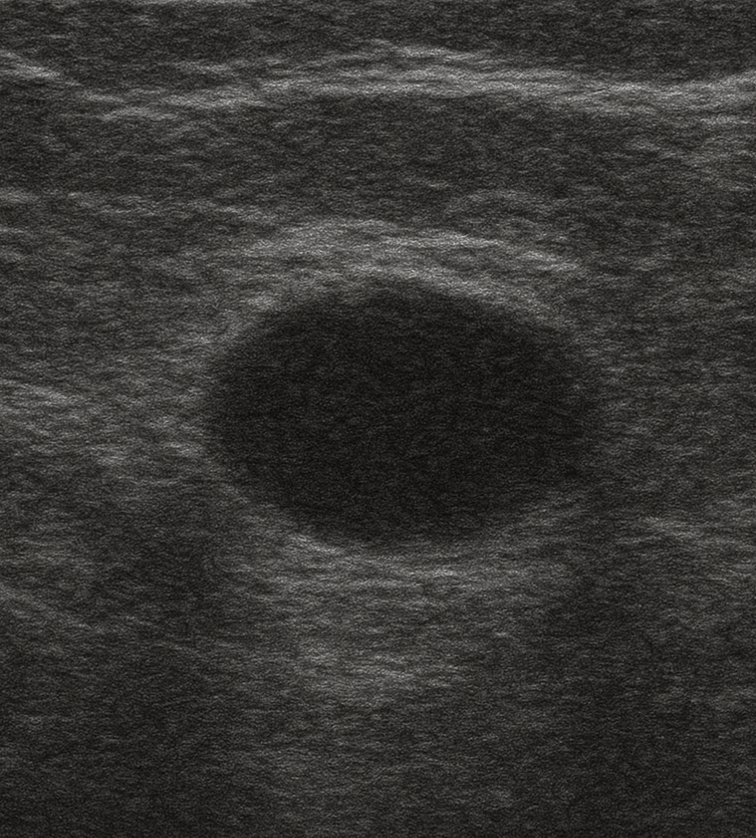}
    }
    \subfloat[DDPM Maligant.]{
        \includegraphics[width=0.3\linewidth, height=0.3\linewidth]{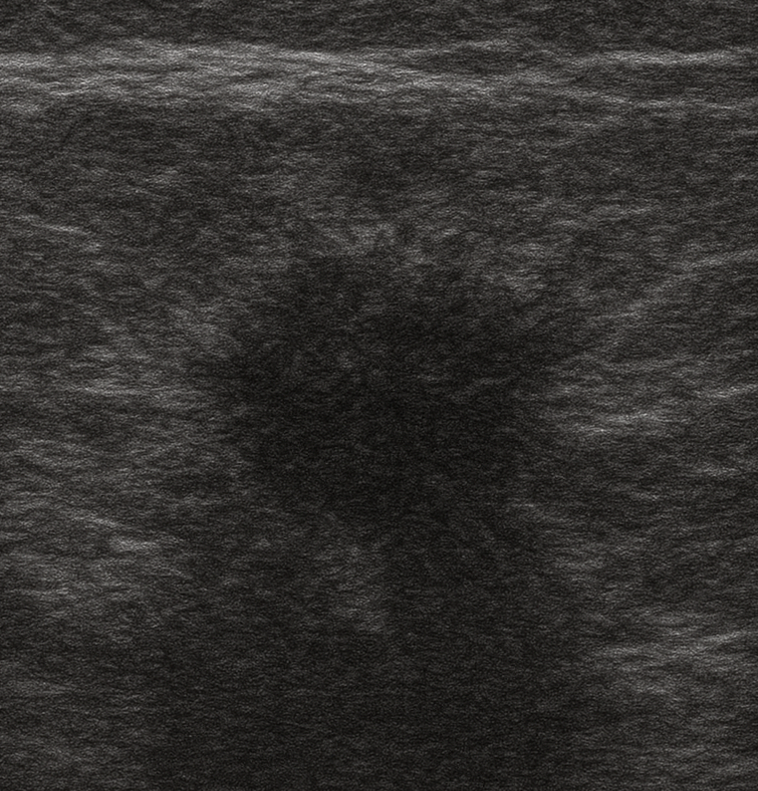}
    }
    \\
    \caption{Samples of synthetic breast ultrasound images. 
    }
    \label{fig: synthetic}
\end{figure}

\begin{table*}[tb]
    \caption{Federated breast ultrasound image classification performance with and without synthetic image augmentation.}
    \centering
    \begin{tabular}{lc|cccc|cccc}
    \toprule
         && \multicolumn{4}{c|}{\bf{Accuracy}} & \multicolumn{4}{c}{\bf{AUC}} \\
        \multirow{-2}{*}{\bf{Method}}&\multirow{-2}{*}{\bf{Generative Model}}& \bf{BUSBRA} & \bf{BUSI} & \bf{UDIAT} & \bf{Average} & \bf{BUSBRA} & \bf{BUSI} & \bf{UDIAT} & \bf{Average} \\
         \midrule
         Centralized Learning&-&0.8507& 0.9000&0.9394& 0.8967&0.9199& 0.9510&0.9463&0.9391 \\
         \midrule
         FedAvg&-&0.8480&0.8923&0.8788&0.8730&0.9191&0.9048&0.9380&0.9206\\
         FedAvg&DCGAN&0.8773&0.8615&0.9394&0.8928&0.9373&0.9164&0.9174&0.9237\\
         FedAvg&DDPM&0.8827&0.8769&0.9394&0.8997&0.9410&0.9231&0.9446&0.9362\\
         \midrule
         FedProx&-&0.8747&0.8923&0.9394&0.9021&0.9318&0.9424&0.9545&0.9429\\
         FedProx&DCGAN&0.8720&0.9077&0.9091&0.8963&0.9454&0.9532&0.9628&0.9538\\
         FedProx&DDPM&0.8880&0.9077&0.9394&0.9117&0.9502&0.9586&0.9635&0.9574\\
         \bottomrule
    \end{tabular}
    \label{tab: results}
\end{table*}

\subsection{Implementation Details}

For synthetic image generation, two class-specific DCGAN models (one for benign cases and one for malignant cases) and a class-conditioned DDPM were trained using the combined training and validation data. After training, each model generated 1,000 synthetic benign and 1,000 synthetic malignant images. These synthetic images were uniformly distributed to all clients for data augmentation during federated training. Representative examples of the generated synthetic images are shown in Fig.~\ref{fig: synthetic}.

For the classification task, we utilized ImageNet-pretrained DenseNet-121~\cite{huang2017densely} as the backbone model. The federated learning setup simulated three clients (BUS-BRA, BUSI, and UDIAT datasets). We adapt FedAvg and FedProx as baseline federated learning algorithms. At each communication round, the global model was distributed to all clients, where each performs local training for one epoch with a batch size of 32 using the AdamW optimizer
. The initial learning rate was set to $0.001$ and decayed by a factor of 10 every 30 epochs. For FedProx, the proximal term coefficient was set to $\mu=0.03$. After local training, models were aggregated at the server using weighted averaging proportional to the number of training samples at each client. To prevent overfitting on synthetic data, the initial learning rate for synthetic image batches was set to $0.0001$ and decayed by a factor of 10 every 30 epochs. In each epoch, we randomly selected 160 synthetic images (5 batches) for training, ensuring that synthetic data comprised approximately 12\% of each client's training data. In the ablation study, we observed that using too many synthetic images can negatively impact model performance, highlighting the importance of careful balancing between real and synthetic data. The entire training process was conducted for 100 communication rounds, and the global model achieving the highest average AUC across validation sets was selected for final evaluation on the test set.

\subsection{Federated Breast Ultrasound Image Classification Results}
Table~\ref{tab: results} presents the quantitative results of the federated breast ultrasound image classification experiments. The introduction of synthetic images during federated training consistently improved performance for both FedAvg and FedProx. With DCGAN-based augmentation, the average AUC for FedAvg increased from 0.9206 to 0.9237, and FedProx improved from 0.9429 to 0.9538. Using the proposed class-conditioned DDPM further enhanced these gains, with FedAvg achieving an average AUC of 0.9362 and FedProx reaching 0.9574. For reference, the centralized learning model trained on the combined dataset achieved an AUC of 0.9391, showing that the proposed federated framework with diffusion-based augmentation can not only close the gap with centralized training but even surpass it in some cases, all while preserving data privacy. These results demonstrate that diffusion-generated synthetic images are more effective than GAN-based ones for enriching client datasets, improving convergence, and boosting classification performance in a federated setting.

\subsection{Impact of A Larger Volume of Training Synthetic images}
We then performed an ablation study to investigate the impact of introducing a larger volume of synthetic images during training. However, we observed that while introducing a moderate number of synthetic images improved both ACC and AUC, excessive synthetic augmentation degraded performance due to overfitting to artificial features. For instance, as Fig.~\ref{fig: ablation} shows, 2000 additional synthetic images from DCGAN in training reduced the average test ACC and AUC of the classification model from 0.8928 to 0.8580 and from 0.9237 to 0.9203, respectively. This performance degradation was likely due to the synthetic images introducing distributional noise, which may have diluted the informative patterns in the real data. This finding highlights the importance of carefully balancing synthetic and real data when using generative augmentation in federated learning. 


\begin{figure}[tb]
    \centering
    \subfloat[Accuracy.]{
    \includegraphics[width=0.8\linewidth]{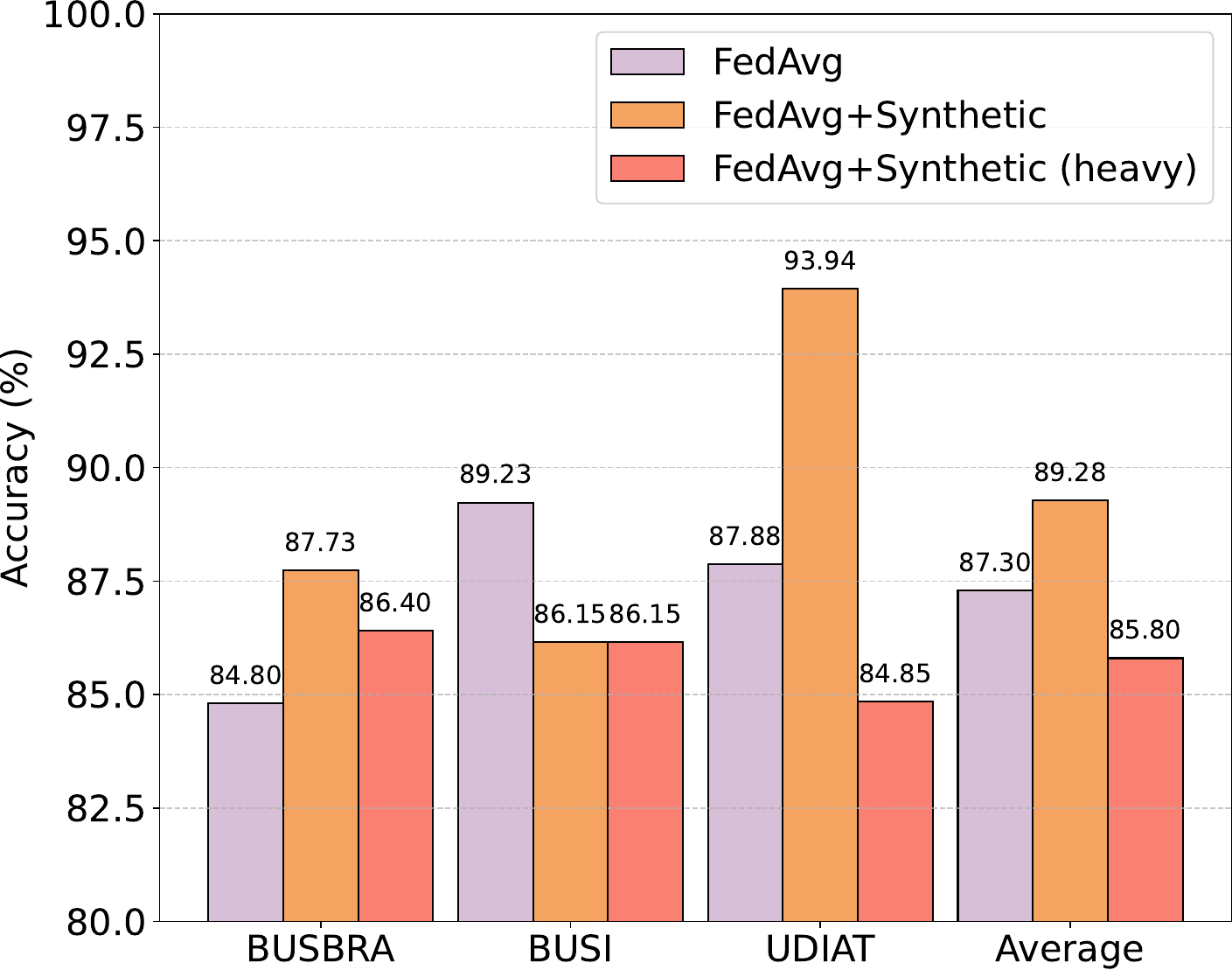}
    }\\
    \subfloat[AUC.]{
    \includegraphics[width=0.8\linewidth]{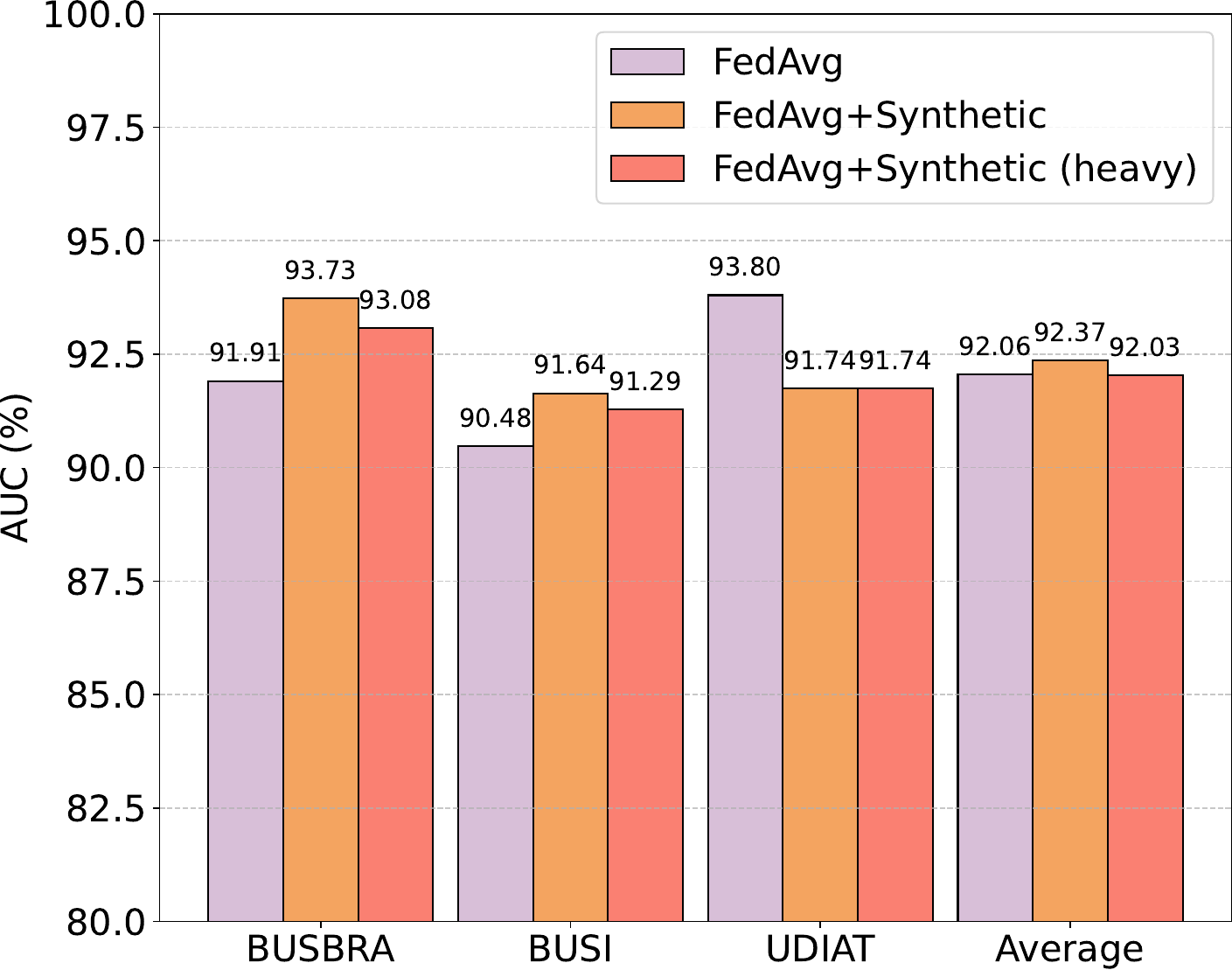}
    }
    \caption{Ablation study results showing the effect of increasing the number of synthetic images during federated training. While moderate augmentation improves performance, excessive synthetic data leads to a decline in both AUC and accuracy, highlighting the importance of balancing real and synthetic samples.}
    \label{fig: ablation}
\end{figure}


\section{Conclusion}
In this work, we developed a federated learning framework for breast ultrasound image classification augmented with synthetic breast ultrasound images. These synthetic images were generated by two class-specific DCGANs or a class-conditioned DDPM. Using three publicly available BUS datasets (BUSI, BUS-BRA, and UDIAT), we simulated a realistic multi-center FL setup and evaluated the impact of generative augmentation. Our results demonstrated that incorporating a suitable number of synthetic images consistently improved model performance, enhancing both FedAvg and FedProx. However, excessive synthetic data reduced accuracy and AUC, highlighting the need to balance synthetic and real samples to avoid over-regularization or domain shift. These findings indicate that generative augmentation can effectively improve federated learning for medical imaging. While our current framework prioritizes efficiency with DCGANs, future work could explore diffusion-based models to further enhance the realism and diversity of synthetic images.

\section{Ethics Statement}
This study utilized the BUS-BRA~\cite{gomez2024bus}, BUSI~\cite{al2020dataset}, and UDIAT~\cite{yap2017automated} datasets, which are publicly available for research purposes. All data were de-identified by the original collectors prior to public release. Consequently, this research involved the secondary analysis of anonymized data and was conducted in accordance with the principles of the Declaration of Helsinki.

\bibliographystyle{IEEEtran}
\bibliography{refs}

\end{document}